\documentclass[twocolumn,prb,amsmath,amssymb,showpacs,
superscriptaddress,floatfix]{revtex4}

\usepackage{graphicx,epsfig}

\DeclareMathOperator{\dt}{det} \DeclareMathOperator{\Tr}{Tr}

\DeclareMathOperator{\Real}{Re}
\begin{document}

\title{Conductance oscillations with magnetic field of a two-dimensional electron gas-superconductor junction}
\author{Nikolai~M. Chtchelkatchev}
\affiliation{L.D.\ Landau Institute for Theoretical Physics, Russian Academy of Sciences, 117940
Moscow, Russia}

\author{Igor~S. Burmistrov}
\affiliation{L.D.\ Landau Institute for Theoretical Physics, Russian Academy of Sciences, 117940
Moscow, Russia}

\begin{abstract}
We find the current voltage characteristics of a 2DEG-S interface
in magnetic field taking into account the surface roughness.
Typically in experiments $L/2R_c\gtrsim 3$, where $L$ is the
surface length and $R_c$ is the cyclotron radius. The conductance
behaves in experiments usually as
$G=g_0+g_1\cos(2\pi\nu+\delta_1)$; higher harmonics,
$g_2\cos(4\pi\nu+\delta_2),\ldots$, are hardly seen. Theories
based on the assumption of the interface perfectness can hardly
describe qualitatively the visibility of the condctance
oscillations and the amplitudes of the harmonics: they predict
$g_1\sim g_2\sim g_3,\ldots$. Our approach with the surface
roughness qualitatively agrees with experiments. It is shown how a
disorder at a 2DEG-S interface suppresses the conductance
oscillations with $\nu$.
\end{abstract}

\pacs{74.80.Fp, 71.70.Di, 73.20.-r, 73.40.-c}

\maketitle

\section{Introduction}

The study of hybrid systems consisting of superconductors (S) in contact with clean 2D normal
metals (2DEG) in magnetic field has attracted considerable interest in recent
years.\cite{review1,review2,review3} The quantum transport in this type of structures can be
investigated in the framework of Andreev refection.\cite{Andreev} When an electron quasiparticle
in a normal metal (N) reflects from the interface of the superconductor (S) into a hole, Cooper
pair transfers into the superconductor. A number of very interesting phenomena based on Andreev
reflection had been studied in the past. For example, if the normal metal is surrounded by
superconductors, so we have a SNS junction, a number of Andreev reflections appear at the NS
interfaces. In equilibrium this leads to Andreev quasiparticle levels in the normal metal that
carry considerable part of the Josephson current; out of the equilibrium, when superconductors are
voltage biased, quasiparticles Andreev reflect about $2\Delta/eV$ times transferring large quanta
of charge from one superconductor to the other. This effect is called Multiple Andreev Reflection
(MAR).\cite{Tinkham}

Effect similar to MAR appears at a long enough N-S interface in magnetic field when the magnetic
field bends quasiparticle trajectories and makes quasiparticles reflect many times from the
superconductor. If phase coherence is maintained interference between electrons and holes can
result in periodic, Aharonov--Bohm-like oscillations in the magnetoresistance. The conductance $G$
of a S--2DEG interface in magnetic field was measured in
experiments.\cite{Takayanagi-Akazaki,Uhlisch,Moore,Eroms,Batov} It showed highly nonmnotonic
dependence with the magnetic field $B$ [large filling factors were considered]; the most
interesting effect was the oscillations of $G$ with the filling factor $\nu$ in a somewhat similar
manner as in Shubnikov-de Gaas effect.

A phenomenological analytical theory of these phenomena based on an ``analogy'' with the
Aaranov-Bohm effect was suggested in Ref.\onlinecite{Asano}. Numerical simulation was made in
Ref.\onlinecite{Takagaki}. It was theoretically shown that the transport along the infinitely long
S/2DEG interface can be described in the framework of electron and hole edge states.\cite{Hoppe}
2DEG-S interfaces investigated in the experiments were not infinitely long, but with the length,
$L$, of the order of few cyclotron orbits, $R_c$, of an electron in 2DEG at the Fermi energy.
Quasiclassical theory of the charge transport through 2DEG-S interface at large filling factors,
arbitrary length of the 2DEG-S interface was suggested in Ref.\onlinecite{ch1}. Most mentioned
above theoretical papers considered the ideal 2DEG-S interface: no roughness. It was
shown\cite{ch1} that when $L\sim 2R_c$, $G(\nu)$ oscillates nearly harmonically with $\nu$, as
$\cos(2\pi\nu)$. When $L\sim 4R_c$ harmonics $\cos(4\pi\nu)$ becomes visible and so on... In
experiments $L\gtrsim 6R_c$, so one would expect good visibility of $\cos(n\pi\nu)$-harmonics in
the conductance, where $n=1,2,\dots$. But if we try to compare theoretically predicted $G(\nu)$
with the experimentally measured one then we find that 1) at $L\gtrsim 6R_c$ only the lowest
harmonic $\cos2\pi\nu$ is seen in the conductance and higher harmonics are absent; 2) the
visibility [amplitude] of the conductance oscillations is much smaller than theories predict. The
reason of this disagreement is probably the roughness of the 2DEG-S interface in experiments and
ideal flatness of this interface in theory.

We try to find in this paper the current voltage characteristics of a 2DEG-S interface in magnetic
field taking into account the surface roughness. Our approach with the surface roughness possibly
helps to make a step towards explanation of the the experimental results. It is shown that the a
disorder at a 2DEG-S interface suppresses high harmonics of the the conductance oscillations with
$\nu$.

We consider a junction consisting of a superconductor, 2DEG  and a normal conductor segments (see
Fig.1). Magnetic field $B$ is applied along  $z$ direction, perpendicular to the plain of 2DEG. It
is supposed that quasiparticle transport is ballistic (the mean free path of an electron
$l_{tr}\gg L$, where $L$ is the length of the 2DEG-S boundary). The current $I$ is supposed to
flow between normal (N) and superconducting (S) terminals (the voltage $V$ is applied between
them).

Following Ref.\onlinecite{Blonder, Takane_Ebisava,Blanter-Buttiker,Datta}, we shall describe the
transport properties of the junction in terms of electron and hole quasiparticle scattering
states, which satisfy Bogoliubov-de Gennes (BdG) equations. Then the current through the 2DEG-S
surface is
\begin{multline}
\label{Takane_Ebisava} I(V)=\frac{e}{h}\int_0^\infty dE \{f_{e}\Tr[\hat 1-R_{ee}+R_{he}]-
\\
f_{h} \Tr[\hat 1-R_{hh}+R_{eh}]\},
\end{multline}
where $$f_{e(h)}=\frac{1}{e^{E\mp eV}+1},$$ $V$ is the voltage of the normal terminal, the energy
$E$ is counted from $\mu$ of the superconductor, $R_{ee}(E,n_o,n_i)$ is the  probability of the
(normal) reflection of an electron with the energy $E$ incident on the superconductor in  the edge
channel with quantum number $n_i$ to an electron going from the superconductor in the channel
$n_o$; the trace is taken in the channel space. Spin degrees of freedom are included into the
channel definition.
\begin{figure}[thb]
\epsfxsize=80mm  \epsffile{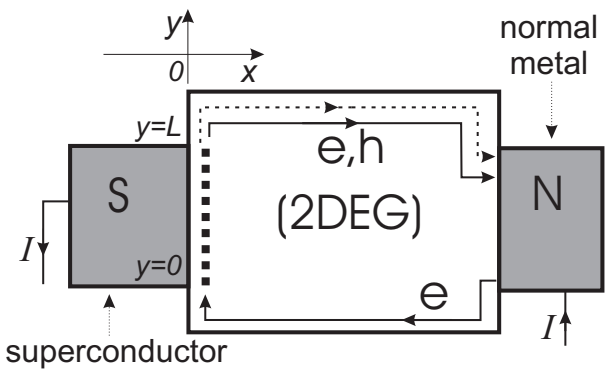} \caption{\label{fig1}The device, which we investigate,
consists of a superconductor, 2DEG and a normal conductor. An electron injected from the normal
conductor in IQH regime goes through an edge state to the superconductor, reflects into a hole and
an electron which return to the normal contact through the other edge states.}
\end{figure}

So the main task of our work is to find the probabilities $R_{ee}$, \textit{etc.} Then we'll be
able to evaluate the current, the conductance, noise and so on.\cite{Blanter-Buttiker} We'll focus
on the situation when If $R_c\lesssim L$. Then quasiparticles reflected from the superconductor
(S) due to normal and Andreev reflection of the electron  return to  S again due to bending of the
trajectories by magnetic field.

From the first glance it may seem that the reflection probabilities $R_{ab}$, $a,b=e(h)$ could be
be found using the ``standard'' approach: by matching the incident and outgoing  quasiparticle
wave functions at $y=0$ and $y=L$ with the linear combinations of the quasiparticle wave functions
at the 2DEG-S boundary corresponding to Andreev edge states.\cite{Hoppe} However this procedure
does not look efficient at large filling factors [experimental parameter range] and especially for
a disordered 2DEG-S interface. Next it is difficult to do the matching in practice because the
Andreev bound states wave functions\cite{Hoppe} and the wave functions of 2DEG edge states are
localized in different domains in $\hat x$ direction. The Andreev bound states wave functions
penetrate inside the superconductor on the length scale of the order of $\xi$,\cite{Hoppe} but
2DEG edge states wave functions of the incident and outgoing electrons do not penetrate inside the
2DEG edges so deep as $\xi$.

At large filling factors the quasiclassical approximation is applicable. We show in this paper
that within the quasiclassical approximation the matching problem can be solved and $R_{ab}$,
$a,b=e(h)$ explicitly evaluated.

An electron (hole) quasiparticle in 2DEG can be viewed in semiclassics as a  beam of
rays\cite{Baranger_Stone,Richter} (in a similar way propagation of light  is described in optics
within eikonal approximation in terms of ray beams\cite{Born}). Trajectories of the quasiparticle
rays can be found from the equations of classical mechanics. In terms of the wave functions this
description means that we somehow make wavepackets from edge states wave functions. Reflection of
an electron from the superconductor is schematically shown in Figs.\ref{fig1}-\ref{fig4}.

\section{Ideally flat 2DEG-S interface}
\begin{figure}[t]
\epsfxsize=85mm \epsffile{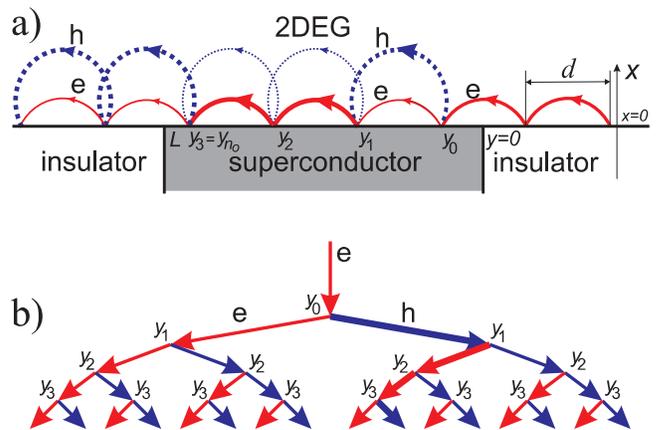}  \caption{\label{fig:fig2} The propagation of quasiparticles
in the quasiclassical approximation can be described in terms of rays. The state of a ray can be
found from the equations of classical mechanics. The figures (a-b) show what happens if an
electron ray from an edge state of 2DEG comes to the superconductor. The electron ray reflects at
$y=y_0$ from the superconductor into electron and hole rays [normal and Andreev reflection]. They
reflect in turn at $y_1$ from the superconductor generating two other electron (red lines) and two
other hole rays (blue lines), and so on. To find the probability, e.g., $R_{he}(E,n_o,n_i)$, it is
necessary to know the sum of the amplitudes of the eight holes that appear  after the last
beam-reflection at $y_3$. Drawing the Fig.\ref{fig:fig2}a we assumed that Andreev
approximation\cite{Andreev} is applicable [see Eq.\ref{eq:cond_Andreev_apr} for conditions] and
the 2DEG-S interface is ideally flat. }
\end{figure}

The transport properties of the ideally flat 2DEG-S interface can be most simply described.  The
edge channels do not mix at such interface. It means in quasiclassical language that electrons
(holes) skip along the 2DEG edge along the same arc-trajectories, like it is shown in
Fig.\ref{fig:fig2}.  Then $n_o=n_i$ and, e.g., $R_{he}(y_0;n_o,n_i)\propto\delta_{n_o,n_i}$.

\begin{figure}[t]
\begin{center}
\includegraphics[height=55mm]{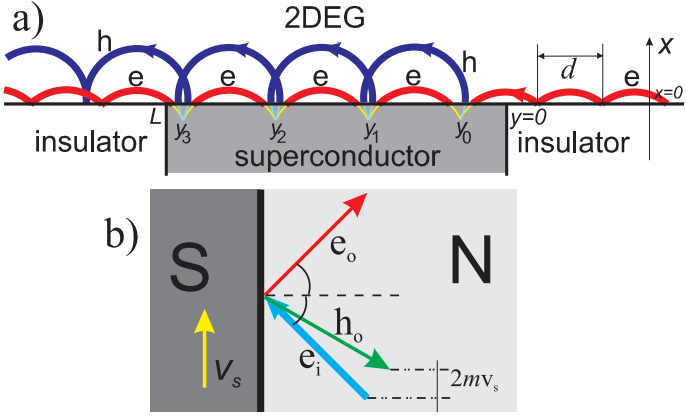}
\caption{When  the conditions of the Andreev approximation are violated we can not use the
assumption that the Andreev-reflected hole velocity is exactly the opposite to the velocity of the
incident at the superconductor electron. Then the quasiparticle rays propagate along the 2DEG-S
interface as scetched in the figure. The orbits are organized as if the scattering occurs not at
the 2DEG-S interface but from the interface in the superconductor lying at some distance  from the
2DEG-S interface. This is so because Andreev reflection couples electron and hole orbits with the
guiding center $x$-coordinates $-\delta\pm X$ [see Ref.\onlinecite{Hoppe}], where $\delta\simeq
l_B^2 mv_s/\hbar$, $l_B=\sqrt{\hbar c/eB}$ and the value of $v_s$ should be taken at the 2DEG-S
interface; for superconductors wider than the London penetration length, $\delta=\lambda_M$.
Fig.\ref{fig3}b illustrates schematically how Andreev and normal quasiparticle reflections occur
from the superconducting interface that carries the supercurrent: $i$ labels the incident
quasiparticle, $o$ --- the reflected ones. } \label{fig3}
\end{center}
\end{figure}

The probability of Andreev reflection can be found as follows for the trajectories shown in
Fig.\ref{fig:fig2}-\ref{fig3}:
\begin{multline}
\label{estimate} R_{he}(y_0;n_o,n_i)=
\\
\delta_{n_o,n_i}\left|e^{i(S_e-\pi/2)}\left\{r_{he}r_{ee}r_{ee}r_{ee}e^{3iS_e-i3\pi/2-i\phi(y_3)}+
\right.\right.
\\
\left.\left.r_{hh}r_{he}r_{ee}r_{ee}e^{iS_h+2iS_e-i\pi/2-i\phi(y_2)}+\ldots\right\}\right|^2,
\end{multline}
where  $r_{ba}$ is the amplitude of reflection of a quasiparticle (ray) $a$ into a quasiparticle
(ray) $b$ from the superconductor; $S_{e(h)}$ is the quasiclassical action of an electron (hole)
taken along the part of the trajectory connecting the adjacent points of reflection; $\pm\pi/2$ is
the Maslov index\cite{Maslov} of the electron trajectory. The phase $\phi(y)$ arises due to the
screening supercurrents. We assume that the superconductor satisfies the description within the
London theory [usual in experiments] then $\phi(y)=\phi(0)+\,\hbar^{-1} \int^y_0 d\tilde{y} 2m
v_s(\tilde y)$,\cite{Svidzinsky} where $v_s$ is the superfluid velocity evaluated at $x=0$ and $m$
is electron mass in the superconductor. We used here the property of London superconductors that
the spatial dependence of the vector potential and $v_s$ are small in the perpendicular direction
to the superconductor edge on the length scale $\xi$ [on which the wave functions of the
scattering electron and hole quasipartciles penetrate in the superconductor].

We introduce the matrix
\begin{gather}
M(y)=\begin{pmatrix}
  r_{\rm ee}\,e^{i(S_e-\pi/2)} & r_{\rm eh}\,e^{i(S_h+\pi/2)+i\phi(y)} \\
  r_{\rm he}\,e^{i(S_e-\pi/2)-i\phi(y)} & r_{\rm hh}\,e^{i(S_h+\pi/2)} \\
\end{pmatrix}
\end{gather}
that contains the amplitudes of Andreev ($r_{\rm he},r_{\rm eh}$) and normal ($r_{\rm ee},r_{\rm
hh}$) quasiparticle reflection from the superconductor at the point $y$.  Then the matrix product
\begin{gather}
S^{(3)}=M(y_3)M(y_2)M(y_1)M(y_0),
\end{gather}
describes the Andreev and normal scattering amplitudes for the case shown in
Fig.\ref{fig:fig2},\ref{fig3}. So,
\begin{gather}
R_{he}(E,y_0;n_i,n_i)=|S_{21}^{(3)}|^2,\\ R_{ee}(E;y_0;n_i,n_i)=|S_{11}^{(3)}|^2.
\end{gather}
If there are n reflections from the 2DEG-S interface then $3\to n-1$.

Below we show that the matrix $S^{(n)}$ can be calculated analytically for any integer $n$ when
the superfluid velocity $v_s(x=0,y)$  is constant so the  phase $\phi(y)$ is a linear function of
$y$. Then the difference, $\phi(y_n)-\phi(y_{n-1})=\delta\phi$, does not depend on $n$ because the
2DEG-S interface is flat and $y_n-y_{n-1}=\ldots=y_1-y_{0}$. The matrix $M(y_n)$ can be written as
\begin{gather}
M(y_n)=\Phi^{\dag}(n)M(y_0)\Phi(n),
\end{gather}
where
\begin{gather}
\Phi(n)\equiv\begin{pmatrix}
  \exp\{-\frac i2n\,\delta\phi\} & 0 \\
  0 & \exp\{\frac i2n\,\delta\phi\} \\
\end{pmatrix}.
\end{gather}
Thus
\begin{gather}
\label{eq:Sn} S^{(n)}=\Phi^\dag(n)(M\Phi)^{n+1}\Phi^\dag,
\end{gather}
where $M=M(y_0)$, $\Phi=\Phi(1)$.

Then [see Appendix 1]
\begin{multline}\label{eq:S}
(M\Phi)^{n+1}=\dt^{(n+1)/2}\cdot
\\
\begin{pmatrix}
  m_{ee}U_{n}(a)- U_{n-1}(a)& m_{eh}U_{n}(a) \\
  m_{he}U_{n}(a) & m_{hh}U_{n}(a)- U_{n-1}(a) \\
\end{pmatrix}\,,
\end{multline}
where $\dt=\{r_{\rm ee}r_{\rm hh}-r_{\rm eh}r_{\rm he}\}\exp[i(S_e+S_h)]$,
\begin{gather}
a=\frac{r_{\rm ee}e^{i(S_e-\pi/2)-\frac i2 \delta\phi}+ r_{\rm hh}e^{i(S_h+\pi/2)+\frac i2
\delta\phi}}{2\sqrt{\dt}},
\\
m_{ee}=\frac{r_{ee}}{\sqrt{\dt}}e^{i(S_e-\pi/2)},
\\
m_{eh}=\frac{r_{eh}}{\sqrt{\dt}}e^{i(S_h+\pi/2)},
\\
m_{he}=\frac{r_{he}}{\sqrt{\dt}}e^{i(S_e-\pi/2)},
\\\label{eq:m22}
m_{hh}=\frac{r_{hh}}{\sqrt{\dt}}e^{i(S_h+\pi/2)}.
\end{gather}

Eqs.\eqref{eq:Sn}-\eqref{eq:m22} give an opportunity to find the probabilities $R_{ab}$, if given
the amplitudes of local reflection, $r_{ab}$. The probabilities $R_{ab}$ depend on the position of
the first reflection from the 2DEG-S interface, $y_0$, that varies in the range $(0,d)$, see
Fig.\ref{fig:fig2}a. The number of reflections, $n$, depends on the choice of $y_0$. So the
solution strategy is to calculate the current using Eq.\ref{Takane_Ebisava} with the probabilities
$R_{ab}$ defined in Eqs.\eqref{eq:Sn}-\eqref{eq:m22} and average the result over $y_0$. The
natural choice for the distribution of $y_0$  is the uniform distribution:
$P_{n_i}[y_0]=\theta[d(n_i)-y_0]/2R_c$. So
\begin{multline}
\label{Takane_Ebisava_f} I(V)=\frac{e}{h}\int_0^\infty dE\sum_{n_i}\int dy_0 P_{n_i}(y_0)
\{[1-R_{ee}+R_{he}]f_{e}-
\\
[1-R_{hh}+R_{eh}]\}f_{h}.
\end{multline}

The action $S_{e}=\int  \mathbf{\tilde k}_{e}\cdot d\mathbf{l}$, where $\tilde k_e$ is the
generalized momentum and the integral is over the quasiparticle trajectory that connects the
adjacent points of reflection from the superconductor-2DEG interface.
\begin{gather}
S_{e}=k_{e} l_e+ \frac{e}{\hbar c} \int \mathbf{A}\cdot d\mathbf{l}=k_e l_e-\frac {|e|}{\hbar
c}\Phi_e,
\\
S_{h}=-k_h l_h+\frac {|e|}{\hbar c}\Phi_h,
\end{gather}
where $k_{e(h)}=\sqrt{2m[\mu_{_{\rm 2DEG}}\pm (E+g\mu_B\sigma B)]/\hbar^2}$, $\sigma=\pm1$, $l$ is
the trajectory length and $\Phi_{e(h)}$ is the absolute value of the magnetic field flux through
the area bounded by the quasiparticle trajectory arc and the 2DEG-S interface. The actions can be
explicitly written in terms of the filling factor $\nu$ and the $y$-component of the quasiparticle
velocity, $v_y^{e(h)}$, at the 2DEG-S interface when $E,g\mu_B\sigma B\ll\mu_{_{\rm 2DEG}}/\nu$:
\begin{gather}
S_{e(h)}=s_{e(h)}\pm\pi(\nu+1/2),
\\\label{eq_s}
s_{e(h)}=2\left(\nu+\frac12\right)\,\left(\arcsin
\Upsilon_{e(h)}-\Upsilon_{e(h)}\sqrt{1-\Upsilon_{e(h)}^2}\right),
\end{gather}
where $\Upsilon_{e(h)}=v_y^{e(h)}/v^{e(h)}$.

Often the Andreev approximation\cite{Andreev} can be used. Then the Andreev-reflected hole
velocity direction may be considered the opposite to the velocity direction of the incident at the
superconductor electron  [see Fig.\ref{fig:fig2}]. The conditions are:
\begin{gather}\label{eq:cond_Andreev_apr}
v_s\ll v_F^{\rm(2DEG)},\quad \max(|eV|,T,g\mu_B B)< \Delta < E_F^{\rm(2DEG)},
\end{gather}
 Then $s_e=s_h$ and
\begin{gather}
S_{e}-S_h=\frac {|e|}{2\hbar c}\Phi=2\pi\left(\nu+\frac12\right),
\end{gather}
where $\Phi$ is the flux through the Larmor ring-trajectory of an electron in magnetic field $B$
at the Fermi shell.

The problem how to evaluate the $r_{ab}$ amplitudes also simplifies within the parameter range,
Eq\eqref{eq:cond_Andreev_apr}. The conditions mean that the magnetic field could be neglected in
the Bogoliubov-de Gennes (BdG) equations [$B$ is already taken into account by the phase $\phi$].
Then $r_{ab}$ can be evaluated according to the BTK theory.\cite{Blonder}

When  the conditions, Eq.\eqref{eq:cond_Andreev_apr}, are violated our quasiclassical transport
picture in terms of quasiparticle rays can be still applied, see Fig.\ref{fig3}. Then the
amplitudes $r_{ab}$ are the solutions of the scattering problem for Bogoliubov--de Gennes
equations:
\begin{gather}
(E-g\mu_B B )u=\left(\frac{[\mathbf{p}+m\mathbf{v}_s]^2}{2m}-\mu\right)u+\Delta\, v,\\
(E-g\mu_B B )v=-\left(\frac{[\mathbf{p}-m\mathbf{v}_s]^2}{2m}-\mu\right)v+\Delta\, u.
\end{gather}
Here $m$, $g$ and $\mu$ should be considered different in S and 2DEG. The spatial distribution of
the superfluid velocity is fixed by the London equation, $\mathrm{rot}\,m
\mathbf{v}_s=-\mathbf{B}\,e/c$.  Usually there is a barrier at the 2DEG-S interface, we did not
write its contribution to BdG explicitly.

It follows from Eq.\eqref{Takane_Ebisava_f} that at zero temperature and voltage the conductance
is:
\begin{gather}
\label{G_Ch} G = \frac{2e^2}{h} \sum_{n_i}\sum_s P_s
\frac{|r_{eh}|^2\sin^2[s\arccos(\sqrt{|r_{ee}|^2}\cos(\Omega))]}{1-|r_{ee}|^2 \cos^2(\Omega)},
\end{gather}
where [we remind again] spin is included at the definition of the channel index,
$\Omega=(S_{e}-S_h-\pi)/2 + \theta -\delta\phi/2$; $\theta=\mathrm{arg}(r_{ee})$ is the phase of
the amplitude of electron -- electron reflection from the superconductor. If the superconductor
characteristic dimensions are larger than $\lambda_M$ -- the Meissner penetration length of the
superconductor then $\delta\phi/2=2\lambda_M k_{\perp}$, where $k_{\perp}=k_{\perp}(n_i)$ is the
perpendicular component the quasiparticle momentum when it reflects from the superconductor. The
function $P_s$ is the probability that the orbit describes $s$ reflections from the surface of the
superconductor; this function originates from the averaging over $y_0$ discussed above. $P_s$ can
be expressed through the maximum number of jumps, $g_m=[L/d]$, over the S-2DEG surface with the
length $L$, where $[\ldots]$ denotes the integer part:
\begin{gather}
\label{P_s} P_{s}=
   \begin{cases}
     \frac{L-g_m d}{d}  & \text{if $s=g_m+1$}, \\
     1-\frac{L-g_m d}{d}& \text{if $s=g_m$}, \\
     0 & \text{otherwise}.
  \end{cases}
\end{gather}

The conductance, Eq.\eqref{G_Ch} [as well as the current, Eq.\eqref{Takane_Ebisava_f}] is an
oscillating function of $\nu$:
\begin{gather}
G(\nu)=\sum_{n=0}^\infty g_n\cos(2\pi\nu n+\delta_n),
\end{gather}
where $g_n$ are the Fourier coefficients and $\delta_n$ -- the ``phase shifts''. When the length
of the interface, $L\lesssim 2R_c$, then the leading contribution to the conductance (current)
gives the zero harmonic; while $2R_c\lesssim L\lesssim 4R_c$ then $G\approx g_0+g_1\cos(2\pi\nu
+\delta_1)$; when $4R_c\lesssim L\lesssim 6R_c$ the second harmonics becomes relevant, and so
on...

\begin{figure}[t]
\begin{center}
\includegraphics[height=65mm]{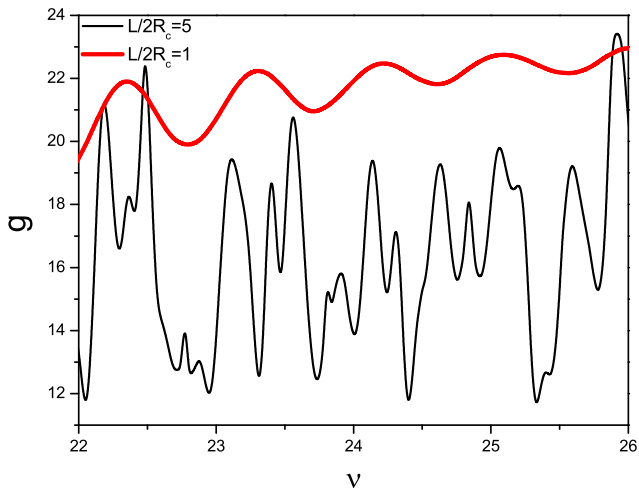}
\caption{The (dimensionless) zero bias conductance oscillations with $\nu$. The parameters are the
following:  $k_F^{\rm(2DEG)}=2\cdot10^6\,\mathrm{cm}^{-1}$, $\delta\phi/4=20$ [corresponds to NbN
film with the width of the order of $100$nm]. The $L=3\,\mu m$ [$L/2R_c\simeq5$ at $\nu=25$] for
the lower curve and $L=0.6\,\mu m$ [$L/2R_c\simeq1$ at $\nu=25$] for the upper (thick) curve. We
neglected the Zeeman splitting (typically small). The 2DEG-S interface scattering amplitudes,
$r_{ab}$, were taken according to the BTK model\cite{Blonder} with $Z=0.6$. The curves were made
using Eq.\eqref{G_Ch}.} \label{fig_g1}
\end{center}
\end{figure}

How the conductance changes with $\nu$ is illustrated in Fig.\ref{fig_g1}. Typically in
experiments $L/2R_c\gtrsim 3$ [thin black curve]. But the conductance behaves in experiments as if
$G=g_0+g_1\cos(2\pi\nu+\delta_1)$;  higher harmonics, $g_3, \ldots$, are not seen. But our theory
based on the assumption of the interface flatness predicts $g_1\sim g_2\sim g_3$ while
$L/2R_c\gtrsim 3$, see Fig.\ref{fig_g1}. The reason of the discrepancy between our theory and the
experiment is the assumption that the 2DEG-S interface is ideally flat. Disorder at the 2DEG-S
interface makes $g_0>g_1>g_2,\ldots$. Below we demonstrate it.


\section{Disordered 2DEG-S interface}

Usually  2DEG-S interface is not ideally flat. The disorder at the interface can be divided at two
classes long range and short range with the respect to the characteristic wavelength, $\lambda_F$,
in 2DEG. Presence of the long range disorder implies that the 2DEG-S interface position fluctuates
around the line $x=0$ at length scales much larger than the $\lambda_F^{\rm(2DEG)}\sim 10^{-6}\mu
m$. Photographs of the experimental setups do not allow to think that 2DEG-S interface bends
strongly from the line $x=0$. So this kind of the disorder is likely not very important.

The short-range disorder includes the fluctuations of the surface at length scales smaller than
$\lambda_F^{\rm(2DEG)}$; impurities, clusters of atoms at the surface due to defects of the
lithography and so on... When, for example an electron ray falls on the disordered 2DEG-S surface
the reflected electron rays go off the surface not at a fixed angle but they may go at any angle
with certain disorder induced probability distribution. The phases that carry the reflected
electron rays going off the surface at different angles may be considered random, so the reflected
electron rays can be considered as incoherent.\cite{footnote_phase} But to any reflected electron
ray an Andreev reflected hole ray is attached that is coherent with the electron. So the
interference of rays [that produces the conductance oscillations] may be not killed completely by
the short range disorder. Below it will be clarified.

\begin{figure}[t]
\begin{center}
\includegraphics[height=100mm]{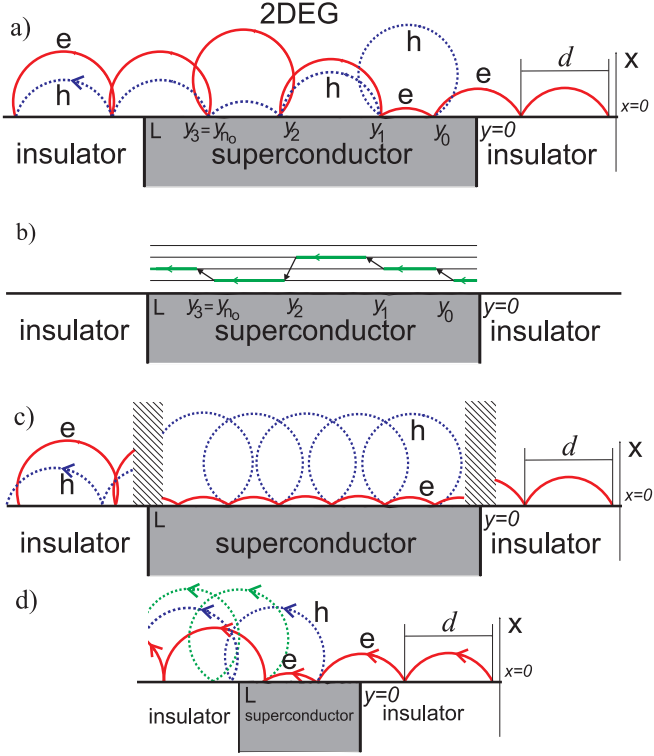}
\caption{Relatively weak short-range disorder at 2DEG-S interface induces transitions between the
Andreev edge states, see Fig.\ref{fig4}b. Andreev edge states in quasiclassics make electron-hole
orbit-arcs with the same beginning and end. The quasiclassical picture of the disorder-induced
transitions is shown in Fig.\ref{fig4}a. Disorder at the edges of the 2DEG-S interface leads to
the orbits depicted in Fig.\ref{fig4}c. Strong disorder destroys Andreev edge states; there are no
oscillations in the conductance because quasiparticles reflected from the strongly disordered
2DEG-S surface are incoherent; orbits are shown in Fig.\ref{fig4}d.} \label{fig4}
\end{center}
\end{figure}

``Weak'' short range disorder at 2DEG-S interface does not destroy the Andreev edge states but it
induces transitions between the edge states, see Fig.\ref{fig4}b. Andreev edge states in
quasiclassics fix electron-hole orbit-arcs with the same beginning and end. The quasiclassical
picture of the disorder-induced transitions is shown in Fig.\ref{fig4}a.

It is shown below how to describe the transport properties of the weakly short-range disordered
2DEG-S interface in magnetic field. We assume that the Andreev approximation conditions,
Eq.\eqref{eq:cond_Andreev_apr}, are fulfilled [general case brings to qualitatively similar
results for the current, it requires same idea of calculations but it is more cumbersome] then
$s_e=s_h$ and the fluctuating quantity is $\delta\phi$. Similarly as before we introduce the
matrix
\begin{gather}
M(y_n)=\Phi^{\dag}(n)M\Phi(n)\,e^{i\sum_{i=0}^n s_i},
\end{gather}
where $s$ is defined in Eq.\eqref{eq_s},
\begin{gather}
\Phi(n)\equiv\begin{pmatrix}
  \exp\{-\frac i2\,\sum_{i=0}^n\delta\phi_i\} & 0 \\
  0 & \exp\{\frac i2\,\sum_{i=0}^n\delta\phi_i\} \\
\end{pmatrix},
\end{gather}
and
\begin{gather}
M=\begin{pmatrix}
  r_{\rm ee}\,e^{i\pi\nu} & r_{\rm eh}\,e^{-i\pi\nu} \\
  r_{\rm he}\,e^{i\pi\nu} & r_{\rm hh}\,e^{-i\pi\nu} \\
\end{pmatrix}.
\end{gather}

As before, for the situation sketched in Fig.\ref{fig4}a,
$S^{(3)}=M(y_3)M(y_2)M(y_1)M(y_0)\,e^{i\sum_{i=0}^3 s_i}$, and,
e.g., $R_{he}=|S_{21}^{(3)}|^2$. It is clear that
$e^{i\sum_{i=0}^3 s_i}$ does not influence on the probabilities
$R_{ab}$ so we'll omit this term below. In general case:
\begin{gather}
\label{eq:Sn_disorder} S^{(n)}=\Phi^\dag(n)[M\Phi_n\ldots M\Phi_1M\Phi_0]\Phi_0^\dag.
\end{gather}
here
\begin{gather}\label{eq:Phi_n}
\Phi_n\equiv\begin{pmatrix}
  \exp\{-\frac i2\delta\phi_n\} & 0 \\
  0 & \exp\{\frac i2\delta\phi_n\} \\
\end{pmatrix},
\end{gather}
where $\delta\phi_n=\phi_n-\phi_{n-1}$ and $\delta\phi_0\equiv 0$.

The probabilities can be found in the following way:
\begin{multline}\label{eq:bubble}
R_{ee}-R_{he}=\frac 1 2\Tr\left\{\sigma_z M\Phi_nM\Phi_{n-1}\ldots
\right.\\\times \left.\Phi_2 M\Phi_1M\sigma_z M^\dag\Phi_1^\dag
M^\dag\Phi_{2}^\dag\ldots \Phi_{n-1}^\dag M^\dag\Phi_{n}^\dag
M^\dag\right\}.
\end{multline}
Here we neglected the term $\Phi^\dag(n)$  that enters Eq.\eqref{eq:Sn_disorder} because it does
not contribute to the probabilities $R_{ab}$. If one wants to find $R_{ee}$ then $\sigma_z$ should
be substituted by $(\sigma_0+\sigma_z)/2$ in the last equation. If $R_{eh}$ is wanted then the
first $\sigma_z$ should be substituted by $(\sigma_0+\sigma_z)/2$, the second -- by
$(\sigma_0-\sigma_z)/2$.

As in the previous section we should average the current over $y_0$. This operation is closely
related to the disorder averaging because shifting $y_0$ we'll make the trajectories go through
different disorder realizations. The phase jumps $\delta\phi_n$ fluctuate due to the disorder.

The disorder average
\begin{gather}\label{eq:Phi}
\langle[\Phi_a]_{ij}[\Phi_b]_{pq}^\dag\rangle=\delta_{ab}\delta_{ij}\delta_{pq}(\delta_{ip}+\Lambda_{ip}),
\\
\Lambda_{ip}=\begin{pmatrix}
  0 & \langle e^{-i\delta\phi_a}\rangle \\
  \langle e^{i\delta\phi_a}\rangle & 0 \\
\end{pmatrix}.
\end{gather}
The averages like $\langle (1-\delta_{ab})e^{-i(\delta\phi_a\pm\delta\phi_b)/2}\rangle=0$ because
quasiparticles acquire a random phase reflecting from the disordered interface as it was mentioned
above.

$R_{ee}-R_{he}$ can be treated perturbatively over small $\Lambda$. The difference $R_{ee}-R_{he}$
can be treated perturbatively over $\Lambda$. It is natural to assume that $\delta\phi_n$ are
gaussian distributed:
\begin{gather}
\langle\delta\phi_a\rangle=\overline{\delta\phi},
\\
\langle\langle\delta\phi_a\,\delta\phi_b\rangle\rangle=2 \eta
\delta_{ab},
\end{gather}
where $\langle\langle\ldots\rangle\rangle$ means the irreducible average. Then
\begin{gather}
\langle e^{i\delta\phi_a}\rangle=
e^{i\,\overline{\delta\phi}}\,e^{-\eta}.
\end{gather}
While $\eta\ll 1$, weak fluctuations,  the results of the previous
section are valid. The opposite case we discuss below.

\begin{figure}[t]
\begin{center}
\includegraphics[height=52mm]{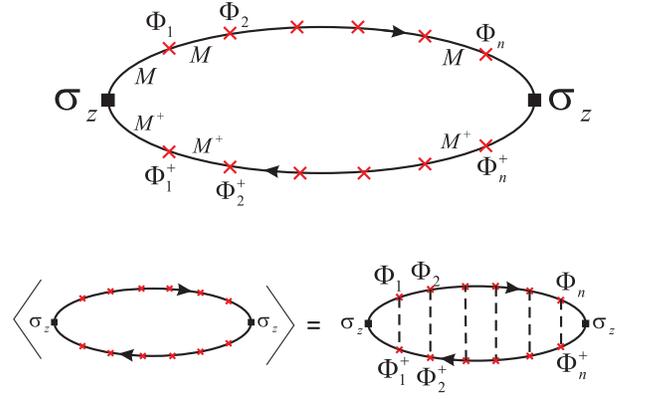}
\caption{Disorder averaging of the transmission probabilities. The loop corresponds to the trace
in Eq.\eqref{eq:bubble}. The $\times$-vertices are $\Phi,\Phi^\dag$, the solid lines represent
$M,M^\dag$, the dashed lines are the $\langle\Phi \Phi^\dag\rangle$ correlators.}
\label{fig:bubble}
\end{center}
\end{figure}

Lets $\eta\gg 1$. In the zero order over $\Lambda$, for $n=3$, the
average probabilities
\begin{multline}
R_{ee}-R_{he}=
\\
\Tr\left\{(\sigma_z)_{i_2i_2}
|M_{i_2i_3}|^2|M_{i_3i_4}|^2|M_{i_4i_5}|^2|M_{i_5i_6}|^2(\sigma_z)_{i_6i_6} \right\}.
\end{multline}
Thus we see that in the ``completely incoherent case'' [zero order
over $\Lambda$] the average probabilities can be found as the
corresponding elements of the matrix $\tilde S^n$:
\begin{gather}
\tilde S^n=\begin{pmatrix}
  |r_{ee}|^2 & |r_{eh}|^2 \\
  |r_{he}|^2 & |r_{hh}|^2 \\
\end{pmatrix}^n,
\end{gather}
for example $R_{ee}=[\tilde S^n]_{11}$. Explicit form of $\tilde
S^n$ can be easily found using the theorem mentioned in the
Appendix~\ref{appendix1}.
\begin{figure}[th]
\begin{center}
\includegraphics[height=48mm]{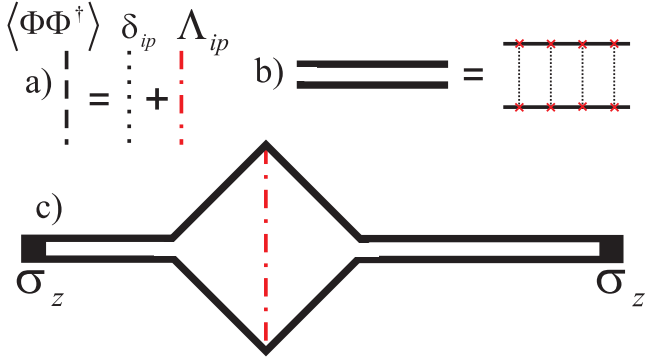}
\caption{Typical contribution to the average of the bubble in the first order over $\Lambda$ is
schematically shown in the figure, Fig.\ref{fig:buble_1order}c. The $\langle\Phi \Phi^\dag\rangle$
line splits into the sum of two: the first one curresponds to the $\delta_{ip}$-term in
Eq.\eqref{eq:Phi} and the second one -- to the $\Lambda$-term, Fig.\ref{fig:buble_1order}a. The
parallel lines show the part of the bubble, Eq.\eqref{eq:bubble}, where the first order of
$\Lambda $ is taken (``diffusion''), it does not carry interference information,
Fig.\ref{fig:buble_1order}b. The diamond means the place where $\Lambda$-line is inserted; this
diamond provides oscillations of the current with $\nu$. One should sum all $n$ diagram like shown
in Fig.\ref{fig:buble_1order} [they differ by the position of the diamond] to find the average of
the bubble in the first order over $\Lambda$.} \label{fig:buble_1order}
\end{center}
\end{figure}
More interesting is the first order expansion of the probabilities
$R_{ab}$ over $\Lambda$. After calculations diagrammatically
illustrated in Fig.\ref{fig:buble_1order} (see
Appendix~\ref{appendix2}) we find for the average probabilities
\begin{multline}\label{eq:A_n}
A_n=R_{ee}-R_{he}=
\\
= R^{n} \left [ 1 + e^{-\eta}\frac{4 (n-1) |r_{ee}|^2
|r_{eh}|^2}{R^2} \cos (2\Omega)\right ]
\end{multline}
where $\Omega = \pi \nu +\theta_{ee} - \overline{\delta\phi}/2$,
$R=|r_{ee}|^2-|r_{eh}|^2$ and $n$ is the number of reflections
from the 2DEG-S interface.  All trajectory-dependent quantities
that enter Eq.\eqref{eq:A_n} should be evaluated for the
trajectory with equal electron and hole arcs
[$\Upsilon_{e(h)}=0$].

\begin{figure}[th]
\begin{center}
\includegraphics[height=40mm]{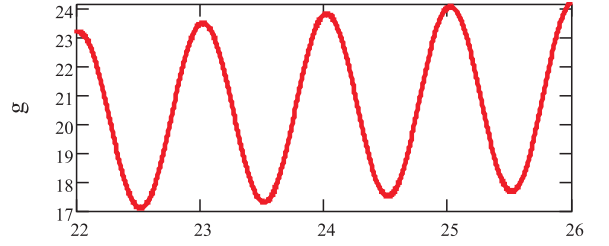}
\caption{The figure shows how the conductance depends on $\nu$
when the interface is disordered for parameters similar to used
for the black curve in Fig.\ref{fig_g1}, $e^{-\eta}\simeq 0.1$.
Now the conductance behavior qualitatively agrees with the
experimental data. } \label{fig6}
\end{center}
\end{figure}

So, finally the zero-bias conductance is
\begin{gather}\label{eq:G-disorder}
G=\frac{4e^2}h \nu\sum_n P_n \,[1-A_n ],
\end{gather}
where $P_s$ is defined in Eq.\eqref{P_s}, but with $d\to 2R_c$. The conductance oscillations
according to Eq.\eqref{eq:G-disorder} are illustrated in Fig.\ref{fig6}.

The result similar to Eqs.\eqref{eq:A_n},\eqref{eq:G-disorder} would be obtain calculating the
influence of the disorder at the edges of 2DEG-S interface, Fig.\ref{fig4}c, on the
magneto-conductance.

\section{Conclusions}
We found in this paper the current voltage characteristics of a 2DEG-S interface in magnetic field
taking into account the surface roughness. Our approach with the surface roughness possibly
removes the contradiction between the theory and the experiment. It is shown that the a disorder
at a 2DEG-S interface suppresses the conductance oscillations with $\nu$. The measurement of the
magneto-conductance of a 2DEG-S boundary is the test for the degree of the boundary roughness.

\section{Acknowledgments}
We would like to thank I.\,E.\,Batov for the detailed discussions
of the experimental data. We thank RFBR 06-02-17519, the Russian
Ministry of Science, the Netherlands Organization for Scientific
Research (NWO), CRDF and Russian Science Support foundations.


\appendix

\section{Appendix 1 \label{appendix1}}

For the sake of a reader's convenience, we present in this
appendix the following theorem.~\cite{Abeles,Born} Let us $Q$ be a
$2\times2$ (complex) matrix with the determinant equal to unity
then


\begin{gather}
    Q^{n+1}=\begin{pmatrix}
  Q_{11}U_{n}(a)- U_{n-1}(a)& Q_{12}U_{n}(a) \\
  Q_{21}U_{n}(a) & Q_{22}U_{n}(a)- U_{n-1}(a) \\
\end{pmatrix},\label{AbelesQ}
\end{gather}
where $a=\Tr Q/2$ and $U_{n}(a)$ is the Chebyshev polynomial of the second kind\cite{Gradshtein}
$$U_{n}(a)=\sin[(n+1)\arccos(a)]/\sqrt{1-a^2}.$$

\section{Disorder averaging\label{appendix2}}

In this appendix we present the details of derivation for
Eqs.~\eqref{eq:A_n}. Let us consider the matrix
\begin{gather}
 \begin{pmatrix}
    A_{n+1} & B_{n+1} \\
    C_{n+1} & D_{n+1}
  \end{pmatrix}
= \left\langle M \Phi_{n+1} \begin{pmatrix}
    A_{n} & B_{n} \\
    C_{n} & D_{n}
  \end{pmatrix}
\Phi_{n+1}^\dag M^\dag\right \rangle
\end{gather}
with
\begin{gather}\label{ABCD0}
\begin{pmatrix}
    A_{0} & B_{0} \\
    C_{0} & D_{0}
  \end{pmatrix}
= M\sigma_z M^\dag .
\end{gather}
Then the quantity $R_{ee}-R_{he}$ in Eq.~\eqref{eq:bubble} is
equal to $A_{n-1}$. It is convenient to introduce a vector $\Psi_n
= (A_n,D_n,B_n,C_n)^T$ such that
\begin{gather}
\Psi_{n+1} = \mathbb{T}\left (e^{-\eta}\right ) \Psi_n,\quad
\mathbb{T}(x)
=
  \begin{pmatrix}
    \tilde{S} & - x U \\
    W & - x V
  \end{pmatrix}
\end{gather}
where the elements $U$, $V$ and $W$ of the transfer-matrix
$\mathbb{T}$ are the following $2\times 2$ matrices
\begin{gather}
 W=\begin{pmatrix}
   r_{ee}r_{he}^\star  & r_{eh}r_{hh}^\star \\
    r_{he}r_{ee}^\star & r_{hh}r_{eh}^\star
  \end{pmatrix},
\\
U =
  \begin{pmatrix}
   r_{ee}r_{eh}^\star e^{i2\pi \nu-i \overline{\delta\phi}} & r_{eh}r_{ee}^\star e^{-i2\pi \nu+i \overline{\delta\phi}} \\
    r_{he}r_{hh}^\star e^{i2\pi \nu-i \overline{\delta\phi}} & r_{hh}r_{he}^\star e^{-i2\pi \nu+i \overline{\delta\phi}}
  \end{pmatrix},\\
   V=\begin{pmatrix}
   r_{ee}r_{hh}^\star e^{i2\pi \nu-i \overline{\delta\phi}} & r_{eh}r_{he}^\star e^{-i2\pi \nu+i \overline{\delta\phi}} \\
    r_{he}r_{eh}^\star e^{i2\pi \nu-i \overline{\delta\phi}} & r_{hh}r_{ee}^\star e^{-i2\pi \nu+i \overline{\delta\phi}}
  \end{pmatrix}.
\end{gather}
By using that $\Psi_0 = \mathbb{T}(0) \Psi_{-1}$ where
$\Psi_{-1}=(1,-1,0,0)^T$, we find
\begin{gather}
\Psi_n = \mathbb{T}^{n}(x) \mathbb{T}(0) \Psi_{-1}. \label{eqs2}
\end{gather}
To the lowest order in $\exp(-\eta)$ we find from Eq.~\eqref{eqs2}
\begin{gather}
A_n = \lim_{x\to 0}\Tr\begin{pmatrix}
    1 & 0 \\
    -1 & 0
  \end{pmatrix}\Bigl [ \tilde{S}^{n+1}
  - 2 e^{-\eta}
\Real \, e^{i(2\pi \nu-\phi)} \notag \\ \hspace{2cm} \times
\frac{\partial}{\partial x} (\tilde{S}+ x Z)^{n}  \Bigr
]\label{f1}
\end{gather}
where
\begin{gather}
Z = \begin{pmatrix}
   r_{ee}^2r_{eh}^\star r_{he}^\star & |r_{eh}|^2r_{ee}^\star r_{hh}^\star \\
    |r_{he}|^2r_{hh}^\star r_{ee} &
    r_{hh}^{\star 2}r_{eh}r_{he}
  \end{pmatrix}.
\end{gather}

\subsection{Zero temperature $T=0$}

At vanishing temperature $T=0$ the matrices $\tilde{S}$ and $Z$
can be simplified drastically
\begin{gather}
\tilde{S} =
  \begin{pmatrix}
    |r_{ee}|^2 & |r_{eh}|^2 \\
    |r_{eh}|^2  & |r_{ee}|^2
  \end{pmatrix},\quad Z = |r_{ee}|^2 |r_{eh}|^2 e^{2i\theta}
\begin{pmatrix}
    -1 & 1 \\
    1  & -1
  \end{pmatrix}.
\end{gather}
With the help of the following identity
\begin{gather}
\begin{pmatrix}
    a & b \\
    b  & a
  \end{pmatrix} = \frac{\sigma_x-\sigma_z}{\sqrt{2}} \Bigl [ a - b\sigma_z \Bigr ]
\frac{\sigma_x-\sigma_z}{\sqrt{2}}
\end{gather}
we obtain Eq.~\eqref{eq:A_n} as
\begin{gather}
A_{n-1} = R^{n} \left [ 1 + 4 (n-1) e^{-\eta} |r_{ee}|^2
|r_{eh}|^2 R^{-2} \cos 2\Omega \right ].
\end{gather}

\subsection{Arbitrary temperature}

At arbitrary temperature (energy) where is no special relations
between $r_{ab}$. Then with the help of Eq.~\eqref{AbelesQ} we
find
\begin{gather}
A_n = A_n^{(0)} - 2 e^{-\eta} \Real \, e^{i(2\pi\nu
-\overline{\delta\phi})} A_n^{(1)}
\end{gather}
where
\begin{gather}
A_n^{(0)} = \textrm{det}_0^{n/2} \left [
\frac{|r_{ee}|^2-|r_{he}|^2}{\sqrt{\textrm{det}_0}}
U_{n-1}(a_0)-U_{n-2}(a_0)\right ]
\end{gather}
and
\begin{gather}
A_n^{(1)} = \textrm{det}_0^{\frac{n-1}{2}} U_{n-1}(a_0) (r_{ee}^2
r_{he}^\star r_{eh}^\star - |r_{eh}|^2 r_{ee} r_{hh}^\star)
 \notag \\
- \textrm{det}_0^{\frac{n}{2}} \left [\alpha a_0
U_{n-2}^\prime(a_0) + \frac{n}{2} \beta U_{n-2}(a_0) \right ]
\notag \\
+ \textrm{det}_0^{\frac{n-1}{2}} \left [ \frac{n-1}{2}
U_{n-1}(a_0)-\alpha a_0 U_{n-2}^\prime(a_0)\right ]
(|r_{ee}|^2-|r_{eh}|^2)
\end{gather}
Here we have introduced the following notations
\begin{gather}
\textrm{det}_0 = |r_{ee}|^2|r_{hh}|^2-|r_{eh}|^2||r_{he}|^2,
\\
a_0 = \frac{|r_{ee}|^2+|r_{hh}|^2}{2\, \sqrt{\textrm{det}_0}}.
\end{gather}
and
\begin{gather}
\beta = \textrm{det}_0^{-1}\Bigl ( |r_{hh}|^2 r_{ee}^2
r_{eh}^\star r_{he}^\star +|r_{ee}|^2 r_{hh}^{\star 2} r_{eh}
r_{he} \notag \\ \hspace{2cm}-2 |r_{eh}|^2 |r_{he}|^2 r_{ee}
r_{hh}^\star\Bigr ), \\
\alpha =\frac{r_{ee}^2 r_{eh}^\star r_{he}^\star + r_{hh}^{\star
2} r_{eh}r_{he}}{|r_{ee}|^2+|r_{hh}|^2} - \frac{\beta}{2}.
\end{gather}


\begin{thebibliography}{99}
\bibitem{footnote_phase} This is usual assumptions that the reflected
electron (hole) rays can be considered as incoherent with the incident ray on a (sligtly)
disorderd interface. This assumption is widely used in nanophysics. Well known
Zaitsev\cite{Zaitsev} boundary conditions for quasiclassical greens functions were derived with
this assumption. With the help of this assumption the so-called ``nondiagonal'' quckly oscillating
components of the greens functions that carried information about the coherence were regarded
quickly fading [if we go off the surface].\cite{Zaitsev,circuittheory}
\end{thebibliography}
\end{document}